\documentclass[12pt]{article}
\usepackage{textcomp}
\usepackage{times}
\usepackage{codarxiv,float,amssymb}
\usepackage[reqno]{amsmath}
\usepackage[]{graphicx}
\floatstyle{ruled}
\newfloat{displaycode}{htp}{alg}[section]
\floatname{displaycode}{Eiffel code}
\usepackage[bookmarks,bookmarksopen,
          pdftitle={"Remarks on Jurdzinski-Lorys"},
          pdfauthor={Colm \'O D\'unlaing}]{hyperref}

\newtheorem{definition}[theorem]{Definition}
\newtheorem{lemma}[theorem]{Lemma}

\newtheorem{corollary}[theorem]{Corollary}

\numberwithin{equation}{section}

\newcommand{\redstar}{{\overset{*}{\rightarrow}}}

\newcommand{\thuecong}{{\overset{*}{\leftrightarrow}}}
\newcommand{\redexes}{\mbox{\rm Redexes}}
\newcommand{\reducts}{\mbox{\rm Reducts}}
\newcommand{\irred}{\mbox{\rm Irred}}
\newcommand{\tm}{\mbox{\rm TM}}

\newcommand{\pal}{\text{\sc Pal}}
\newcommand{\cent}{\mbox{\textcent}}
\DeclareGraphicsExtensions{.eps}
\title{Remarks on Jurdzi\'nski and Lory\'s' proof that
palindromes are not a Church-Rosser language}
\author{Colm \'O D\'unlaing\\
Natalie Schluter\thanks{e-mail: odunlain@maths.tcd.ie,
natschluter@maths.tcd.ie.
Mathematics dept website: http://www.maths.tcd.ie.}\\
{\em Mathematics, Trinity College, Dublin 2, Ireland}}
\begin{document}
\maketitle
\begin{abstract}
In 2002 Jurdzi\'nski and Lory\'s settled a long-standing
conjecture that palindromes are not a Church-Rosser language.
Their proof required a sophisticated theory about computation
graphs of 2-stack automata.  We present their proof in terms
of 1-tape Turing machines.

We also provide an alternative proof of Buntrock and Otto's
result that the
set of bitstrings $\{x: (\forall y) x\not= y^2\}$,
which is context-free, is not Church-Rosser.
\end{abstract}

\section{Introduction}
In the 1970s, Nivat [\ref{nivat}] began the study of
languages defined by Thue systems: see also [\ref{cn},\ref{berstel}].
Book [\ref{book}] continued the study of Church-Rosser Thue
systems, and the theory has been much extended since then
[\ref{bo},\ref{mcn}].

We follow the definitions of length-reducing Thue systems,
etcetera, in [\ref{bo}].
A Thue system $S$ is {\em Church-Rosser} if whenever
$$ u \thuecong_S v,$$ there exists a string $w$
such that $u\redstar w$ and $v\redstar w$.
Equivalently,
every congruence class contains exactly one irreducible string.
The redexes, reducts, and irreducible strings, with respect
to $S$, are denoted $\redexes(S)$, $\reducts(S)$,
and $\irred(S)$.

%\begin{lemma}
%\label{lem: any old reduction}
%If $S$ and $T$ are Thue systems, where $S$ is Church-Rosser,
%$T \subseteq \thuecong_S$,
%and $\irred(T) = \irred(S)$, then $T$ is Church-Rosser
%and equivalent to $S$.
%\end{lemma}
%
%{\bf Proof.}
%Suppose $[x_1]_S = [x_2]_S$. Since
%$\thuecong_T \subseteq \thuecong_S$, $[x_i]_T \subseteq [x_i]_S$,
%$i=1,2$. Let $x_0$ be
%the unique irreducible string (modulo $S$) in
%$[x_1]_S=[x_2]_S$.
%
%Now, $[x_1]_T \cap \irred(T) \not= \emptyset$,
%and 
%$$[x_1]_T \cap \irred(T) \subseteq [x_1]_S \cap \irred (S)
%= \{x_0\},$$ so $[x_1]_T \cap \irred (T) = \{x_0\}$.
%Therefore $T$ is Church-Rosser.
%Also, $[x_2]_T \cap \irred{T} = \{x_0\}$, so $[x_1]_T = [x_2]_T$:
%$T$ is equivalent to $S$. {\bf Q.E.D.}\medskip

$\pal$ denotes the set
of (bitstring) {\em palindromes}: those bitstrings
which read the same backwards as forwards, namely,

$$ \pal = \{ x \in \{0,1\}^*:~ x^R = x \} $$ where $x^R$ is the
reversal of $x$.

Church-Rosser languages will be described below.  They
are a surprisingly powerful generalisation of
congruential languages, which are finite
unions of congruence classes of a finite Church-Rosser Thue
system.

In [\ref{berstel}] it is shown that $\pal$
is not a congruential language. This is proved
by contradiction. Otherwise, by definition, $\pal$
is a finite union of congruence classes of a
Thue system $T$.\footnote{That $T$ is Church-Rosser does not affect
the argument.}

However,
the linguistic congruence
$\equiv_{\pal}$ is the identity relation.  It is defined by
$$x \equiv_\pal y \iff \text{(def.)}~~
(\forall u,v) (uxv \in \pal \iff uyv \in \pal).$$ If
$x$ and $y$ are different bitstrings, suppose without loss
of generality that $|x| \leq |y|$ and $y$ ends in $1$. Then
$$ \lambda x 0^{|y|} y^R $$ is not
a palindrome but
$ \lambda y 0^{|y|} y^R $ is.  Thus $x \equiv_\pal y \iff x=y$.
But $\thuecong_T$ would be a refinement of $\equiv_\pal$, so
$\thuecong_T$ would be the identity relation, and $\pal$, being infinite, would
not be a finite union of congruence classes modulo $T$.

\begin{definition}
\label{def: church-rosser language}
A language $L$ is {\em Church-Rosser} {\rm [\ref{mno}]} if
there exists a Church-Rosser
Thue system $S$ and  strings $t_1$, $t_2$, and $t_3$,  such that
$$ \{t_1\}\cdot L \cdot \{t_2\} = [t_3]_S .$$ We
assume without loss of generality that $t_3$ is irreducible,
so $x\in L$ if and only if $t_1 x t_2 \redstar_S t_3$

We only consider languages $L\subseteq\{0,1\}^*$.
The alphabet of $S$ may include $\{0,1\}$ properly.
\end{definition}

Church-Rosser languages were introduced in 1984 by
Narendran [\ref{naren}], and
studied in [\ref{mno}]
by McNaughton, Narendran, and Otto.

Book [\ref{book}] had shown that if $S$ is a Church-Rosser
Thue system then reduction (modulo $S$) could be
executed in linear time on a 2-stack automaton.
Therefore Church-Rosser languages can be
recognised on a ``shrinking'' deterministic 2-stack automaton.
Two papers by
Buntrock and Otto [\ref{buno}] and
Niemann and Otto [\ref{no}]
together showed that such automata characterise
the Church-Rosser languages.

An early conjecture by
McNaughton, Narendran, and Otto  [\ref{mno}] was that
the language of bitstring palindromes is {\em not} Church-Rosser.
This conjecture
remained open until it was proved by
Jurdzi\'nski and Lory\'s in 2002 [\ref{jl}].

Jurdzi\'nski and Lory\'s'
proof (see [\ref{jl07}]) is difficult, requiring
a complex theory of computation graphs for
two-stack automata.  In this note we propose
a simplified proof based on 1-tape Turing machines.

\section{1-tape reduction machine}
Given a Church-Rosser Thue system $S$, we exhibit
a 1-tape Turing machine $\tm$ implementing reduction modulo
$S$ in a systematic way.  While Book's 2-stack machine is more
efficient [\ref{bo},\ref{hartmanis}],
the advantage of studying
reductions on a 1-tape Turing machine is that
blanks are steadily accumulated, allowing us to
see where information has been lost.

`Turing machine' will mean a deterministic machine with
quintuple instructions and 2-way infinite tape,
although the worktape used will be only slightly
longer than the input string.  An instruction (quintuple) has
the form

\centerline{
current state, current symbol, new symbol, head movement,
new state}

\noindent
where the head movement is 1 square left or right (the read/write
head moves at every step).

Given a language $L$ such that
$$ t_1 \cdot L \cdot t_2 = [t_3]_S,$$ on input $x$
the machine $\tm$ converts the tape contents
to $t_1 x t_2 $, reduces it, and compares the result
to $t_3$.
Let $\Sigma$
be the smallest alphabet such that

\centerline{$\Sigma^*$
contains $\redexes(S)$, $\reducts(S)$,
$\{t_1,t_2,t_3\},$ and $L$.}

%Book's reduction machine [\ref{bo}] was designed
%to run in linear time, and
%1-tape Turing machines hardly ever do so
%([\ref{hartmanis}]).

The machine $\tm$ executes reductions systematically.
If a string $z$ is reducible, then it has
a leftmost redex, i.e.,
it can be written as
$wut$ where $u$ is a redex and no proper prefix of $wu$ is
reducible.

The set of such strings $wu$ is regular and one can
easily describe a DFA $D$ which recognises this set,
and has the property that when it accepts
$wu$, one such redex $u$, and
hence a rule $u\to v$, is determined uniquely by
its accepting state.
Ties are broken arbitrarily.

Let $K$ be the set of states of $D$.

The worktape alphabet of $\tm$ consists of

\begin{itemize}
\item
$\Sigma$, a new blank symbol $B$, and left
and right {\em sentinel characters} $\cent$ and $\$$.
\item
{\em Compound symbols} $[a,k]$ where $a \in \Sigma\cup\{B\}$
and $k\in K$ (the states of $D$).
\end{itemize}

The {\em blank symbols} are
$$ B ~\text{and}~\{ [B,k]:~ k \in K\}.$$

\numpara
\label{par: hom h}
Write $h$ for the following homomorphism.
\begin{eqnarray*}
h(z) = \begin{cases}
z \quad \text{if} ~ z \in \Sigma,\\
a \quad \text{if} ~ z = [a,k],~a\in \Sigma, ~k\in K,\\
\lambda\quad\text{otherwise}
\end{cases}
\end{eqnarray*}

Let $k_0$ be the initial state of $D$ and
$\delta$ the transition function for $D$.
We
extend $\delta$ to $K\times (\Sigma\cup\{B\})$:
$$ \delta (k,B) = k,\qquad k\in K .$$

A string $[a_1,k_1][a_2,k_2] \ldots [a_n,k_n]$ of
compound symbols is {\em historical} if for all $j\leq n$,
$$ k_j = \delta^*(k_0, a_1 a_2 \ldots a_j).$$
Obviously,
$$ k_{j+1} = \delta (k_j, a_{j+1}),\quad 0 \leq j \leq n-1.$$

\begin{definition}
\label{def: initial redex}
The string $\cent t_1 x t_2 \$ $ (including endmarkers) is
called the {\em initial redex} on input $x$.
\end{definition}

The machine $\tm$ creates the intial redex, then
reduces as often as needed.

\begin{itemize}
\item
Its configurations are represented in the form $\alpha q \beta$ where
$\alpha\beta$ are the tape contents, including
$\cent$ on the left and $\$$ on the right, $\beta\not= \lambda$
(so $\$$ is the rightmost symbol in $\beta$),
$q$ is the current state, and the machine is scanning
the first symbol of $\beta$.

\item
Except for the
sentinel characters, all symbols in $\beta$
are in $\Sigma\cup\{B\}$ and all symbols in
$\alpha$ are compound symbols, and $\alpha$ is historical.

\item
After $\cent t_1$ and $t_2\$ $ have been added to the input,
$h(\alpha)$ is always irreducible and
$h(\alpha\beta)$ is always a reduct of $t_1 x t_2$ (except temporarily
during REDUCE phases).

\item
First, $\tm$ moves to the right, appending $t_2\$$ to $x$.
Then it moves to the left, prefixing $\cent t_1$ to $x t_2\$$:
the tape contents are now the initial redex $\cent t_1 x t_2 \$ $,
and the current symbol is $\cent$.
It enters a SHIFT phase.

For the rest of this description $\alpha q\beta$ denotes
the current configuration, and $a$ is the current symbol.
\item
In a SHIFT phase, if $\beta = \$$,
then $\tm$ enters its final phase, described below.

Let $k'=k_0$ if $\alpha = \lambda$ or $\alpha=\cent$,
otherwise let $k'$ be the state of $D$ occurring
in the rightmost symbol $[a',k']$ in $\alpha$.
$\tm$ can remember $k'$.

If $a=\cent$ then $\tm$ moves right.

If $a = B$ then $\tm$ overwrites the current
square with $[B,k']$ and moves right.

Otherwise $a \in \Sigma$:
let $k =\delta(k',a)$.
If $k$ is not an accepting state of $D$ then
$\tm$ overwrites the current square with
$[a,k]$ and moves right.

Otherwise, $k$ is an accepting state of $D$,
and the string $h(\alpha) a$ ends in
a redex $u$, so there exists a rule
$u\to v$ associated with $k$.  $\tm$ enters
a REDUCE phase.

\item
In a REDUCE phase, $h(\alpha ) a$ ends with
a redex $u$, and $\tm$ can select a unique rule
$u\to v$ to be applied.
$\tm$ moves left, overwriting the rightmost
$|v|$ symbols of $\alpha a$ with $v$, extending
leftwards with blank symbols $B$, until the
square holding the leftmost symbol $\ell$ of $u$
(or rather, a compound symbol $[\ell,k]$) is overwritten,
moves one square further left, scanning either
$[\ell',k']$ or $\cent$ (in which case let $k'=k_0$),
moves right, writes $[B,k']$, and enters a SHIFT phase.

\item
In the final phase, $\beta = \$$ and the tape contents are
$\cent\alpha\$,$ and $h(\alpha)$ is
irreducible.  $\tm$ scans leftwards to determine
whether or not $h(\alpha) = t_3$, and
halts.

\item
Let $L$ be the maximum length of all redexes.
In a REDUCE phase at most $L+1$ nonblank symbols are scanned,
and the number of blank symbols increases by at
least 1.

\item
There is one left-sweep at the beginning when $\tm$
writes $\cent t_1$.  Thereafter every left-sweep
is a reduction and increases the number of blank symbols.
\end{itemize}

\numpara
\label{par: blank symbols dont affect outcome}
{\bf Blank symbols do not affect the outcome.}
It is very important that the blank symbol $B$
carries no information, and once a square becomes
blank it remains blank (compound symbols
$[B,k]$ are also considered blank).  If one
were to insert extra blank squares at any time,
provided that $B$ is inserted right of the
current square and the appropriate symbols
$[B,k]$ are inserted left of the current square,
the same reductions would be performed.

\section{Kolmogorov complexity}
\newcommand{\utm}{\mbox{\rm UTM}}

We use the following definition of
the Kolmogorov complexity $K(w)$ of a bitstring $w$.

Let the entire family
of 1-tape Turing machines (transducers,
converting bitstring inputs to bitstring outputs)
be encoded as bitstrings
and a Universal Turing machine $\utm$ be given.  The
encoding of Turing machines should have the property
that if $y$ encodes a Turing machine then no
proper prefix of $y$ does.  In that case,
for any bitstring $x$ there exists at most
one possible factorisation $yz$ of $x$ such
that $y$ encodes a Turing machine, call it $T_y$.

On input $x$,
$\utm$ tests whether
$x$ has a prefix $y$ encoding a Turing machine.
If not, it loops.  Otherwise it
simulates $T_y$ on input
$z$ where $x=yz$, either looping or computing
$T_y(z)$.

For any bitstring $w$, there exists a shortest string
$ yz $
such that $T_y$ computes $w$ on input $z$.

The {\em Kolmogorov complexity} $K(w)$ is the length
of this shortest string.
%It is known that  a constant $C$ exists such
%that $K(w) \leq |w|+C$ for all $w$:
%choosing for example
%a string $y_0$ encoding a Turing machine which does nothing,
%on input $w$ its output is $w$, and
%$\utm$ computes $w$ on input $y_0w$.
%Therefore $K(w) \leq |w| + |y_0|$ for all $w$.

Given bitstrings $w,y,z$ such that $w = T_y(z)$
we say that $yz$
{\em encodes} $w$, or, by abuse of language,
say that $z$ {\em encodes} $w$,
and call $z$ the {\em code} and $y$ the {\em decoder}.

If $K(w) \geq |w|$ then we call $w$ {\em hard.}  The lemma
below is a fundamental result
but very easily proved.

\begin{lemma}
\label{lem: hard strings exist}
For any $m\in \IN$, there exists a hard string $w$ of length $m$.
\end{lemma}

{\bf Proof.} There are $2^m-1$ strings of length $<m$, so there
are at most $2^m-1$ (decoder,code) pairs $yz$ such
that $|yz|<m$.  Hence there exists at least one string
$w$ of length $m$ not encoded by any of them.
{\bf Q.E.D.}\medskip

\section{Crossing sequences and information loss on
a 1-tape reduction machine}
On input $x$,
the reduction machine $\tm$ first creates the initial redex
$$ \cent t_1 x t_2 \$ .$$ Suppose that the initial redex has length $n$
and that the tape squares are labelled $1$ to $n$, beginning with
the $\cent$. The square initially scanned has index
$ |\cent t_1| + 1.$

In discussing crossing sequences, it helps to consider the
`points separating' adjacent squares. There are
{\em crossing points} between squares $i$ and $i+1$
for $0 \leq i \leq n$. During its computation, $\tm$
occasionally moves from square $i$ to $i+1$, or vice-versa;
it is said to {\em cross} the $i$-th crossing point.
This is possible only if $1 \leq i \leq n-1$.

\begin{definition}
\label{def: u,v crossing point}
Given a factorisation $\cent t_1 x t_2 \$ =uv$ of the
initial redex, the $u,v$-crossing point is the crossing
point indexed $|u|$.  Or given a factorisation $x=uv$ of
the input string, the $u,v$-crossing point is the crossing
point indexed $|\cent t_1 u|$.
\end{definition}

During a computation of $\tm$, for $1 \leq i \leq n-1$
a {\em crossing sequence} develops at the $i$-th crossing point,
as follows.

If $i > |\cent t_1|$ then the first crossing is from
left to right,when $\tm$ attaches
$t_2\$$ to the input string, and the second
is from right to left before $\tm$ attaches $\cent t_1$
to the input string.  If $1 \leq i \leq|\cent t_1|$
then the first crossing is from right to left.
The next square scanned is the $i+1$-st if crossing
from left to right, otherwise it is the $i$-th.

Let $p_1$ be the state immediately after the first crossing:
the next square scanned is scanned in state $p_1$.
After that, the crossing point is crossed in the
opposite direction, or possibly never.  Let
$p_2$ be the state immediately after the second crossing, if any.
Then let $p_3$ be the state immediately after the third crossing, and so on.

The initial direction of movement across
the crossing-point is leftwards (resp., rightwards)
if the crossing point is left (resp., right) of the initial
square.  Accordingly crossing sequences begin with
a single bit $s$ indicating whether the point is left (0)
or right (1) of the initial square.

The sequence
$$ s, p_1, p_2, \ldots, p_k $$ is called the {\em crossing sequence}
at the $i$-th crossing point, where $s$ is $0$
if the $i$-th crossing point is left of the initial square, otherwise
$1$.

The bit $s$ is called the {\em leading bit} in the crossing sequence.

The number $k$ is the {\em height} of the crossing sequence.
It ignores $s$: a crossing sequence of height 0 is a single bit.

Because of the repeated introduction of blanks, we can
establish a notion of when significant information
has been lost.  We call a string $y$
{\em depleted} when the number of nonblank
symbols falls below a certain threshold.
(The threshold $1/7$ will be good enough.)

\begin{definition}
\label{def: depleted}
Suppose that the alphabet of the Thue system realised by $\tm$
contains $A$ symbols.  Suppose
$\alpha$ is fixed, $0 < \alpha < 1$.  Let
$$\beta =\frac{\alpha}{\lceil \log_2 A \rceil}.$$
Let $j_1 < j_2$ be two crossing points.  The tape contents between
$j_1$ and $j_2$ are {\em depleted} (at time $t$) if the string
$y'$ between these crossing points satisfies
$$ |h(y')| \leq \beta (j_2 - j_1).
$$  If $y$ is a distinguished
substring of an input string then we say that $y$ is
{\em depleted} at time $t$ if the initial redex $\cent t_1 x t_2 \$ = uyv$ and
the tape contents become depleted as described, where
$j_1 = |u|$ and $j_2 = |uy|$. In this case, 
$$ |h(y')| \leq \beta |y|.$$
\end{definition}

The constant $\beta$ is introduced because
it is the {\em bit-length} of
$h(y')$ which matters,
that is, the length of a bit-string encoding $h(y')$.
The depletion lemma guarantees that $h(y')$ has
bit-length $\leq \alpha |y|$.

\begin{lemma}
\label{lem: depletion lemma}
{\bf (Depletion Lemma).}
There exist constants
$H$ and $d$ such that during any computation
of $\tm$, if two crossing points are at least
$d$ squares apart and
the height of all crossing sequences at and between
them is at least $H$, then
the string between these points
is depleted.
\end{lemma}

{\bf Proof.}
Let $L$ be the maximum length of all redexes.
 Suppose that at crossing point $j$,
and at time $t$, the crossing sequence has
height $H$ or greater.  This includes $\lfloor H/2 \rfloor$
right-to-left movements.  The first may be when
the string $\cent t_1$ is attached to the input,
and another may be the last move in a
reduce phase, when $\tm$ scans the $j$-th square,
which contains $[a,k']$, say, to ascertain
the state $k'$ of $D$.
However,
at that time the $j+1$-st square goes from nonblank
to blank, so it happens at most once.
Apart from these two exceptions, every right-to-left
movement across the $j$-th crossing point is
during a REDUCE phase and produces
more blanks to the left of that point.  This
happens at least $\lfloor H/2 - 2\rfloor$ times
up to time $t$.

Consider a section of at most $K=\lfloor H/2 - 2\rfloor + L - 1$
squares ending at the $j$-th square.
So long as the section includes $L$ or more
nonblank squares, all of these REDUCE phases increase
the number of blanks in the section.
By time $t$ the section contains at most $L-1$
nonblank squares.

Now suppose that the stated threshold holds at
all crossing points from the $(j-\ell)$-th to the $j$-th inclusive,
where $\ell \geq d$.  Subdivide the tape between these
points into sections
of length $K$ plus one leftmost section of length
between $0$ and $K-1$. This subdivision produces
$\lceil\ell/K\rceil$  sections.
By time $t$, the
number of nonblank squares between these crossing points is at most
$$ \frac{(L-1)(\ell + 1)}{ K }.$$  $K$ depends directly on
$H$.  Choose $H$ large enough so that
$$ \frac{L-1}{K} < \beta:\quad K > \frac{L-1}{\beta}.$$ Choose
$$ d = 
\lceil \frac{1}{\frac{\beta K}{L-1} - 1}\rceil.$$ Then
for all $\ell \geq d$,
$$ \frac{(L-1)(\ell + 1)}{ K }\leq \beta \ell,$$ as
required. {\bf Q.E.D.}\medskip

\section{Cut-and-paste methods}
\label{sect: cut and paste}
\def\labelitemii{$\diamond$}
\newcommand{\next}{N}
\newcommand{\nextstate}{{\mbox{\rm Nustate}}}
%\newcommand{\quint}{{\mbox{\rm quintuple}}}
%
%A typical `crossing-sequence argument' is that if the computation
%of a machine $M$ on input $x$ produces duplicate crossing sequences
%at two crossing points, and $M$ accepts $x$, then that part
%of the input string between the two crossing points can be deleted
%yielding a shorter string $x'$ also accepted by $M$.
%These arguments
%apply to the crossing sequences determined
%when a computation runs to completion.
%We need to discuss partial computations (which lose information).

In a
`cut and paste' method, given an input string $x$,
one replaces a substring $v$ of $x$ with another
string $v'$, so $x=uvw$ is changed to a string $x' = uv'w$.
Given that the crossing sequences around $v$ and $v'$
are compatible, the computations on $x$ and $x'$ should be similar.

We consider partial computations of $M$,
where $M$ is a 1-tape Turing machine.
By `partial' is meant that they begin at initial configurations
but do not necessarily end in halting computations.
Associated with every partial computation
is the list of crossing sequences generated by
the computation.

Recall that a crossing sequence is a sequence of the form
$$ s, p_1, \ldots, p_k $$ where $s$ is a single bit
and $p_1,\ldots, p_k$ are states of $M$.  The leading bit
is always given, but if $k=0$ then the sequence is considered empty.

In this section we assume that the squares are indexed
so the first square scanned has index $1$.

\numpara
\label{par: alternating list}
Given an input string $x$,
there is a unique computation (possibly infinite) on input $x$.
Suppose the initial tape contents are presented as
$a_{K}\ldots a_N$, where $K \leq 1$
and $N \geq |x|$ and $x = a_1 \ldots a_{|x|}$;
the other $a_i$ are blank.
Assume that in any partial computation under consideration,
only squares indexed between $K$ and $N$ are
scanned, perhaps not all of them.

\numpara
\label{par: specification}
Now suppose that we are given an alternating list
of crossing sequences $c_i$ and symbols $a_i$,
\begin{equation}
\label{given data}
c_{K-1} , a_K, c_K, \ldots, a_N, c_N,
\end{equation} where
the leading bit in $c_i$ is $0$ if $i < 1$ and $1$ if $i\geq 1$,
and the input string $x$ is $a_1\ldots a_{|x|}$.

Also, those $i$ such that $c_i$ is nonempty
form a contiguous (possibly empty) interval,
and $c_K$ and $c_N$ are empty, with leading
bits $0$ and $1$ respectively.

\numpara
\label{par: full verification}
{\bf Full verification.}
Given this data, it is easy to trace the computation
on input $x$ and produce a sequence of sextuples
$$ i_{r-1}~ p_{r-1}~ a_{r-1} ~ a_r~ \mu_r~ p_r,
\quad r = 1,2,\ldots $$ giving the square scanned
and the quintuple applied at the first, second, \ldots steps.
At the same time the procedure can check the state $p_r$ against
the relevant crossing sequence ($p_0 = q_0$ is not checked).

This can be done by maintaining
the index of the current square, the current state,
and arrays $A_i,~K\leq i \leq N$ and $I_i, K-1 \leq i \leq N$.
The array $A_i$ gives the current tape contents, and
$I_i$ gives the number of states currently cancelled from
$c_i$. The procedure is simple and we omit the details.

The procedure should continue until either

\begin{itemize}
\item
it reaches a halting configuration,
\item
it attempts to check $p_r$ against a state in some $c_i$
where $I_i$ has reached the height of $c_i$, meaning
that all states in $c_i$ have been `cancelled,' or
\item
it checks $p_r$ against a state in some $c_i$ and
discovers a mismatch.
\end{itemize}

In the first two cases, if all states in all the $c_i$
have been cancelled, it reports `consistent,' else
it reports `inconsistent.'  In the third case, it
reports `inconsistent.'

\numpara
\label{par: local verification}
{\bf Local verification.}
Next let us fix some $k, ~ K\leq k \leq N$, and consider
how this procedure affects the $k$-th square: the relevant
data and variables are
$$ k, c_{k-1}, I_{k-1}, q, A_k, c_k, I_k.$$  Let us
suppose, omitting some simple variants,
that $k\geq 2$, so the square is first entered
from the left.  When the square is first entered,
$q$ has just been cancelled from $c_{k-1}$  and $A_k=a_k$,
and a quintuple $q a_k a' \mu q'$ applies, say.
$A_k:= a',$ $q:=q'$, and the next square scanned is $k\pm 1$
depending on $\mu$: $q'$ is cancelled from $c_k$ or $c_{k-1}$
as appropriate, and the next time the square is entered,
$q$ is taken from $c_{k}$ or $c_{k-1}$.  The procedure
continues until there is a mismatch
or it attempts to check $q'$ against $c_{k-1}$ or $c_k$
when all of it has already been cancelled.  At this
point, if there is a mismatch, or not both these
crossing sequences have been fully cancelled, it reports
`inconsistent,' else it reports `consistent.'  Let
us call this procedure a {\em local verification} at the
$k$-th square.

\begin{definition}
\label{def: consecutive triple}
\label{def: compatible}
Given the data (\ref{given data}), i.e.,
$
c_{K-1}, a_K, c_k, \ldots, a_N, c_N, 
$ a {\em consecutive triple} is a triple
$c_{k-1},a_k,c_k$ where $K\leq k \leq N$.

The consecutive triple
$$ c_{k-1}, a_k, c_{k} $$ is
{\em compatible} if the local verification at the $k$-th square
reports `consistent.'
\end{definition}

\begin{theorem}
\label{thm: nasc for consistency}
The data (\ref{given data}) is consistent with
a partial computation on input $x$ if and only
if for each $k$ between $K$ and $N$ the consecutive triple
$$ c_{k-1}, a_k, c_k $$ is compatible.  In this case
the local verification at $k$ also computes the
contents of the $k$-th square at the end of the
partial computation, and identifies the unique square
at which the partial computation ends.
\end{theorem}

{\bf Proof.}
If the data in (\ref{given data}) is consistent
with a partial computation on input $x$,  the local
verification at every square will have the same effect as
the full verification and report `consistent,'
so $c_{k-1}, a_k, c_k$ are compatible
and the final value of $A_k$ will be the same as
in the full verification.

Granted that the data is consistent, the
unique $k$-th square at which the partial computation
ends is easily determined from $k, c_{k-1}, a_k, c_k$
by checking the final head-movement across
the $k-1$st and $k$th crossing points.

Otherwise, the full verification would
report inconsistency.  Suppose it terminates
at the $k$-th square.
Up to this point, its actions at the $k$-th square
are the same as the local verification procedure on that square, so the local
verification at $k$ will terminate and report inconsistency for
the same reason, and $c_{k-1}, a_k, c_k$ are incompatible.
{\bf Q.E.D.}\medskip

The Jurdzi\'nski-Lory\'s proof uses a kind of pumping lemma
and a kind of splicing lemma.  The pumping lemma is

\begin{corollary}
\label{cor: pumping}
{\bf (Pumping Lemma).}
Suppose that $x$ is an input string and $x=uvw$
where $v\not=\lambda$ and in some partial
computation on input $x$, the $u,vw$-crossing
sequence equals the $uv,w$-crossing sequence.
Explicitly, suppose the data
\begin{equation}
\label{eq: pumping data 1}
c_{K-1}, a_K, c_k, \ldots, a_N, c_N, 
\end{equation}
describes a
partial computation on input $x$. Write
$x = a_1 \ldots a_n$ and $v = a_{i+1} \ldots a_j$.
Let $x' = uw= a_1 \ldots a_i a_{j+1} \ldots a_n$.

Then $i<j,$ $c_i$ and $c_j$ are the $u,vw$- and $uv,w$-crossing
sequences respectively, and
\begin{equation}
\label{eq: pumping data 2}
c_{K-1}, a_K, c_K, \ldots, a_{i-1}, c_i, a_j, c_{j+1}, \ldots, a_N, c_N
\end{equation} is
produced by a partial computation on input $x'$.

Furthermore, if $a'_K \ldots a'_N$ are the tape contents
at the end of the first partial computation, then
$$ a'_k \ldots a'_i a'_{j+1} \ldots a'_N$$ are the
contents at the end of the second.
\end{corollary}

{\bf Proof.}  From Theorem \ref{thm: nasc for consistency},
all triples $c_{k-1}, a_k, c_k$ from the list in Equation
(\ref{eq: pumping data 1}) are compatible.  Since
$c_i = c_j$, the same goes for the list in Equation
(\ref{eq: pumping data 2}), so they are produced by
a partial computation on input $x'$.  The remark about
the final tape contents also holds because they
can be calculated by the local verification. {\bf Q.E.D.}\medskip

The other cut-and-paste result is restricted to
our reduction machine $\tm$.
Recall
(Paragraph \ref{par: hom h})
that $h$ is a homomorphism which erases blank symbols,
and a blank symbol may differ from the specific
blank $B$.

\begin{definition}
\label{def: residue}
Let $\tm$ be a reduction machine
with initial redex  $\cent t_1 x t_2 \$ = uvw$ and suppose that
a computation is executed up to a time $T$.
Let $c_1$ be the $u,vw$-crossing sequence at that point,
and $c_2$ the $uv,w$-crossing sequence, and suppose
that $z$ is the tape contents between these crossing points
at time $T$ (i.e., $z$ is the string occupying
squares $|u|+1$ to $|uv|$ at time $T$).

If at time $T$, the square being scanned is one of these
squares, write $v=\alpha\beta$ where this square
is the first in $\beta$ and let $\ell = |h(\alpha)| + 1$;
otherwise let $\ell=0$.

Let $q$ be the state at time $T$.

Then the data
$$ |v|, c_1, h(z), c_2, \ell, q $$ is called a {\em residue} or
$(u,v,w)$-residue (at time $T$).
\end{definition}

The residue is associated with a distinguished substring $v$ of
the initial redex.  It includes $|v|$, $q$, and $\ell$,
to simplify the `splicing lemma'
(\ref{lem: splicing}) below.

\begin{lemma}
\label{lem: grand residue}
Suppose $x_1$ and $x_2$ are input strings,
and there exist times $T_1$ and $T_2$
such that the $\lambda, \cent t_1 x_1 t_2 \$,\lambda $-residue
at time $T_1$ and the
the $\lambda, \cent t_1 x_2 t_2 \$,\lambda $-residue
at time $T_2$ are the same.  Then $x_1$ and $x_2$
possess the same irreducible reduct, so $\tm$
accepts $x_1$ iff it accepts $x_2$.
\end{lemma}

{\bf Proof.} The respective initial redexes lead
to configurations
at times $T_1$ and $T_2$ which are the same except
for occurrences of blank symbols,  which don't
affect the outcome of the computations
(Paragraph \ref{par: blank symbols dont affect outcome}).
{\bf Q.E.D.}\medskip

\begin{lemma}
\label{lem: splicing}
{\bf (splicing lemma).}
Let $\tm$ be a reduction machine.
Given two computations, with inputs factorised as
$uvw$ and $u'v'w'$,
suppose that at some time $t$ in the first
computation, and another time $t'$ in the second,
the residue of $v$ in the first
coincides with the residue of $v'$ in the second.
Then $uvw$ and $uv'w$ possess the same irreducible
reduct, so $\tm$ either accepts or rejects both
strings.
\end{lemma}

{\bf Proof.}
Suppose the common residue is
$ |v|, c_1, h(z), c_2, \ell, q $.
Associated with the first computation suppose we have the
data
\begin{equation}
\label{eq: presplice 1}
 c_{K-1}, a_K, \ldots, a_N, c_N,
\end{equation} $x = a_1\ldots a_n$,
and $v = a_i \ldots a_j$.  Similarly, for the second, we have the data
\begin{equation}
\label{eq: presplice 2}
c'_{K'-1}, b_{K'}, \ldots, b_{N'}, c'_{N'},
\end{equation}
$x' = b_1\ldots b_{n'}$,
and $v' = b_{i'} \ldots b_{j'}$.  We are given
that $c_i = c'_{i'}$ and $c_j = c'_{j'}$.
By
Theorem \ref{thm: nasc for consistency}, each consecutive
triple in both lists of data is compatible.  Corresponding
to the input $x' = uv'w$ we have the list
\begin{equation}
\label{eq: postsplice}
c_{K-1}, a_K, \ldots, a_i, c_i,
b_{i'+1}, c_{i'+1}, \ldots, b_{j'}, c_{j'},
a_{j+1}, \ldots a_N, c_N,
\end{equation} and each consecutive triple
in this list is compatible.  Therefore
by Theorem \ref{thm: nasc for consistency},
there is a partial
computation on input $x'$ which produces the crossing
sequences (\ref{eq: postsplice}).

The residues include the lengths of $v$ and $v'$, so
$v$ and $v'$ have the same length.

The tape squares where these partial computations end
are determined by the local verifications (Theorem
\ref{thm: nasc for consistency}).  If $\ell=0$ then
the first computation ends outside the range of $v$,
so the third computation ends outside the range of $v'$,
at the same square according to the local verifications.
Therefore

\vspace*{6pt}
($*$)\hspace*{1cm}
\begin{minipage}{5in}
at the end of the third computation, the
$(\lambda,\cent t_1 uv'wt_2\$, \lambda)$-residue
is the same as the $(\lambda, \cent t_1 uvwt_2\$, \lambda)$-residue at the end of the
first computation.
\end{minipage}
\hfil
\vspace*{6pt}

\noindent
From Lemma \ref{lem: grand residue},
$uvw$ and $uv'w$ have the same irreducible reduct.

Let $z$ and $z'$ be
the string in the squares originally occupied by $v$
and $v'$ in the first two computations.

If $\ell>0$ then the first and second computations
end at positions $k$ and $k'$, say, within the ranges
of $v$ and $v'$ respectively.  Factorise $z$ as $\alpha\beta$
where $|\alpha|=k$, and $z$ as $\alpha'\beta'$ where $|\alpha'|=k'$.
Then from the residue, $h(\alpha) = h(\alpha')$ and $h(\beta)=h(\beta')$.
Again we reach the conclusion ($*$), so $uvw$ and
$uv'w$ have the same irreducible reduct. {\bf Q.E.D.}\medskip

\section{Jurdzi\'nski and Lory\'s' proof}
Given a 1-tape reduction machine accepting
all bitstring palindromes, in particular it accepts
all palindromes of the form
$$ (ww^R)^{2i+1}$$ where $ww^R$ is hard.  Jurdzi\'nski and
Lory\'s [\ref{jl},\ref{jl07}]
showed that no deterministic 2-stack automaton can recognise this set,
and their arguments can be applied unchanged
to the 1-tape reduction machine $\tm$.

The string
$$ \cent t_1~ ww^R~ \ldots ~ww^R ~ t_2 \$ $$ can
be viewed as $2i+3$ blocks indexed from
$0$ to $2i+2$.
The middle block
is indexed $i+1$. Block $0$  is $\cent t_1$ and block $2i+2$
is $t_2\$$, and for $1 \leq j \leq 2i+1$, the
$j$-th block is the $j$-th occurrence of
$ww^R$. Blocks 0 and $2i+2$
are the outer blocks, and the others are inner blocks.
We suppose that the machine $\tm$
recognises the set of palindromes and derive
a contradiction.

The crucial lemma is the Middle Block Lemma,
\ref{lem: middle block lemma}, below.  A parameter
$H$ will be fixed according to the Depletion Lemma above;
in fact the depletion threshold $\alpha = 1/7$ 
will be good
enough.
 We first establish a pumping result, because
it effects the choice of constants in the Middle Block Lemma.

Define $Q$ as the smallest integer such that
$\tm$ has fewer than $2^Q$ states.  All states can be
represented as $Q$-bit patterns, and there is an extra
pattern to represent `no state,' used for padding.
Then every crossing sequence of height $\leq H$
can be encoded as a string of $QH+1$ bits.

\begin{lemma}
\label{lem: pumping effect}
{\bf (pumping effect).}
Let $H$ be fixed and $Q$ defined as above, and let $w$
be a hard string of length $m$.
Given input $x = (ww^R)^{2i+1}$ where $i > 8m\times 2^{QH+1}$,
suppose at a certain
time $t$  in the computation, within each inner block to
the left of the middle block there is at least one crossing sequence
of height $\leq H$.

Then $x$ can be factorised as $u_1 u_2 u_3$, so that the
shorter string $x' = u_1 u_3$ is
also of the form $(ww^R)^{2i'+1}$,
and has the property described in Corollary \ref{cor: pumping},
i.e., at some time $t'$ in the computation on input $x'$,
the crossing sequences are the same as corresponding
crossing sequences in the computation on $x$ at time $t$, and
the tape contents agree outside
the region originally occupied by $u_2$.
\end{lemma}

{\bf Sketch proof.} For each $j$, $1 \leq j \leq i$, choose a
crossing point $k_j$ in the $j$-th block where the
crossing sequence has height $\leq H$.  A crossing point
belongs to a block if it is between the crossing points
bounding the block, or coincides with one of them.
Perhaps some crossing points are counted twice, but
no crossing point is counted more than twice, and
therefore there are more than $4m \times 2^{QH+1}$ crossing points
chosen.
The residues $k_j \bmod 4m$ fall into $4m$ classes
and therefore there exists an $r$, $0 \leq  r < 4m$,
such that the set
$$ \{ j:~ k_j \equiv r \mod 4m\}$$ contains more
than $2^{QH+1}$ indices.  This gives more than $2^{QH+1}$
crossing points where the crossing sequences at time
$t$ have height $\leq H$.  There are at most
$2^{QH+1}$ such sequences, so the same sequence must
occur at two crossing points, call them $k_1 $
and $k_2$, where $4m$ divides $k_2 - k_1$.
These crossing points are in the region of tape
originally occupied by the input string.

Factorise $x$ as $u_1 u_2 u_3$ where
$|\cent t_1 u_1|=k_1$ and $|u_2|=k_2-k_1$.  Since
the $u_1,u_2u_3$- and $u_1u_2,u_3$-crossing sequences match, this
factorisation has the properties described in
Corollary \ref{cor: pumping}.  Because
$x$ is an odd power of $ww^R$ and $|u_2|$  is a multiple
of $4m = 2|ww^R|$, $u_1 u_3$ is also an odd power of $ww^R$,
so $u_1 u_3 = (ww^R)^{2i'+1}$ as asserted. {\bf Q.E.D.}\medskip

\numpara
\label{para: note one exceptional block}
{\bf Note.} The above lemma will be combined with
Lemma \ref{lem: middle block lemma} to derive a contradiction.
According to the Middle Block Lemma, if $m$ is sufficiently large
then the middle block is the first to be depleted
on input $x=(ww^R)^{2i+1}$.  That is, the middle block has
reached depletion level and no other block has.
Consider the string $x'=(ww^R)^{2i'+1}$.  It was formed
as follows: in the original computation, the tape was divided into
regions $A,B,C$, and the region $B$ was deleted.  Also,
the middle block is entirely in the region $C$.
In the second computation, the tape has regions $A'$ and $C'$
corresponding to $A$ and $C$.
The block corresponding to the middle block is entirely in
the region $C'$.
All blocks in $A'$ and $C'$ correspond to blocks
in $A$ and $C$.  There may be one other block in $x'$ to consider,
namely, a block straddling $A'$ and $C'$, which does not
correspond to a block in the first computation.
This will be considered again in the proof of the main
result.

\numpara
\label{par: prefix encoding}
{\bf Prefix encoding of numbers.}
We need to encode numbers such as $i$ as bitstrings so that
no encoding is a proper prefix of another encoding.  This
is easily done.
Given a positive integer $r$, first represent it as
a binary number $s$ with leading bit $1$. Let $q$ be the
homomorphism $0 \mapsto 00, 1 \mapsto 01$.  Then $r$ is
represented as
$$ q(s) 11.$$
Also, $0$ can be represented as $11$.
This encoding has the prefix property, and uses
fewer than $ 4 + 2\log_2 (r+1)$ bits.

In applying the Depletion Lemma, we take the
depletion level $\alpha$ to be $1/7$.  Recall that 
$\beta = \alpha / \lceil \log_2 A \rceil$ where $A$
is the size of the Thue system alphabet.

\begin{lemma}
\label{lem: middle block lemma}
{\bf (Middle block lemma.)}
Given input $x = (ww^R)^{2i+1}$, 
where $i \leq 9m \times 2^{QH+1}$,
\footnote{With this
bound on $i$, Lemma \protect\ref{lem: pumping effect}
can be used later.}
let the
computation continue until
some inner block is depleted, but only one,
at time $t$, say.  Then if $m = |w|$
is large enough, and the depletion
level is $1/7$, the block must be the middle block.
\end{lemma}

{\bf Proof.}
Suppose the $j$-th block is the first inner block to
become depleted, and $j\not= i+1$.  For clarity
we suppose $j<i+1$.  We consider the three
blocks $j-1,j,j+1$ together.

The case $j=1$ should be treated separately.  Assume
$j>1$ so we have three consecutive inner blocks.
By the depletion lemma (\ref{lem: depletion lemma}),
there exists a crossing point in the ($j-1$)st
block where the crossing sequence has height $\leq H$.
Choose the rightmost such crossing point within
the ($j-1$)st block, and let its index be $j_1$.
Here
$|\cent t_1| + (j-2)\times 2m \leq j_1 \leq |\cent t_1| + (j-1)\times(2m)$.
Similarly let $j_2$
index the leftmost crossing sequence, in the ($j+1$)st
block, whose height is $\leq H$.

Consider the following data.
\begin{equation}
\label{encoding depleted blocks}
m, i, j, j_1, c_1, j_2, c_2, h(y'),\ell,q
\end{equation} where $y'$ is the string
between crossing points indexed $j_1$ and $j_2$
at time $t$, and $c_1$ and $c_2$ are the crossing
sequences at that time at those points. Also,
$\ell$ indicates the relative position of the
and $q$ is the state reached, as given in a
residue (Definition \ref{def: residue}).

Note that
$j_2 - j_1 \leq 6m$, so by the depletion
lemma (\ref{lem: depletion lemma}),
$h(y')$ can be encoded as a bitstring of length $\leq 6m/7$.
The crossing sequences can be encoded as bitstrings of
length $QH+1$, and  numbers $m,i,$ etcetera, are $O(m)$.
The numbers can be encoded as discussed in Paragraph
\ref{par: prefix encoding} above, allowing all the data
to be encoded uniquely in a bitstring $z$ of length
$$ |z| \leq \frac{6m}{7} + O(\log m).$$

It is straightforward to consider in turn
every string $w'$ of length $m$,
and determine whether, on input $(w'w'^R)^{2i+1}$,
the $j$-th block is the  first to become depleted, and
if so, whether the residue matches the given information.
If there is only one such string $w'$, then $w'=w$, so we have
a way to generate $w$ from the given information.
Suppose that $T_y$ is a Turing machine constructing $w$ from $z$.
Then
$ yz $ encodes $w$.  If $m$ is sufficiently large, then
$|yz| < m,$ contradicting the fact that $w$ is hard.

Therefore there exists another
string $w'$ of length $m$ which is consistent with the
information stored in $z$. Then there exist factorisations
$$ \cent t_1 (ww^R)^{2i+1} t_2 \$ = tvu $$ and
$$ \cent t_1 (w'w'^R)^{2i+1} t_2 \$ = t'v'u'$$ where
$|t|= |t'|, |v| = |v'|, $ and $|u| = |u'|$, and at corresponding
points in the computations the $t,u,v$-residue
matches the $t',u',v'$-residue.  Then $\tm$ accepts
$tv'u$ (Lemma \ref{lem: splicing}).
However, in the string $tv'u$, the $j$-th block $w'w'^R$ in $v'$ has
its mirror image in $u$, which is of the form
$ww^R$, so $tv'u$ is not a palindrome, a contradiction.

The analysis is much the same if the depleted block
is indexed $1$, since block indexed zero
is the same for all input strings.  We conclude that
the middle block is the first to be depleted. {\bf Q.E.D.}\medskip

\begin{theorem}
{\bf (Jurdzi\'nski-Lory\'s.)}
$\pal$ is not a Church-Rosser language.
\end{theorem}

{\bf Proof.} Otherwise there is a reduction machine
$\tm$ as described, and an input string
$$ x = (ww^R)^{2i+1}, \quad i = 9m\times 2^{QH+1}.$$ By
Lemma \ref{lem: middle block lemma}, at some time
during the computation on input $x$,
the middle block becomes depleted, but no other block
is depleted.
By Lemma \ref{lem: pumping effect}, the tape can be
divided into regions $A,B,C$, where $C$ contains the
middle block, and
there exists a shorter
string $x'=(ww^R)^{2i'+1}$ obtained by deleting
$B$. Correspondingly, the tape with input $x'$
is divided into regions $A'$ and $C'$.
According to the lemma, there exists a
time $t'$ in the computation on input $x'$
where the crossing sequences and tape contents
in $A$ and $C$ at time $t$ correspond exactly
to those in $A'$ and $C'$ at time $t'$.  In
the original computation (at time $t$), only
the middle block is depleted. Since all blocks
in the second computation at time $t'$, except perhaps one block
straddling $A'$ and $C'$, are the same as in
the first at time $t$, in the latter computation
at least one block is depleted and at most two.
One corresponds to the original middle block  and
is in region $C'$, to the left of centre.
The other straddles $A'$ and $C'$ and is also to the left
of centre.  One of these blocks is the first
to be depleted in the second computation,
contradicting Lemma
\ref{lem: middle block lemma} for $x'$. {\bf Q.E.D.}

\section{Application to non-squares}
It is relatively easy to prove a result
of Buntrock and Otto's [\ref{buno}] that
the set

$$ L=\{ x \in \{0,1\}^*:~ (\forall y) x\not= y^2\}$$ is not
a CRL.
Here is a sketch proof.

Consider bitstrings of the form
$$ w^4 $$ where $w$ is hard. The
string $\cent w^4 \$$ reduces to
some string $\alpha$ which causes
$w^4$ to be rejected, since $w^4\notin L$.  Consider the first block (occurrence
of $w$)
to be depleted in the computation.  Repeating the arguments of this paper,
whatever is the first block to be depleted,
the data
$$
m, i, j, j_1, c_1, j_2, c_2, h(y'),\ell,q
$$ does not determine $w$ uniquely if $w$
is hard.  There is another string $w'$ of the same length which
can replace one occurrence of $w$, where by Lemma \ref{lem: splicing},
the altered string reduces to $\alpha$ and is rejected,
whereas it belongs to $L$.\qed

\section{Acknowledgements}
The authors are grateful to Friedrich Otto for
helpful comments.

\section{References}
\label{references} % section
\begin{enumerate}
\item
\label{berstel}
Jean Berstel (1977).
Congruences plus que parfaites et langages alg\'ebriques.
{\em Seminaire\hfil\break
d'Informatique Th\'eorique,} Institut
de Programmation, 123--147.
\item
\label{book}
Ronald V. Book (1982).
Confluent and other types of Thue systems.
{\em J. ACM \bf 29,} 171--182.
\item
\label{bo}
Ronald V. Book and Friedrich Otto (1993).
{\em String-Rewriting Systems.}
Springer-Verlag Texts and Monographs on Computer Science.
\item
\label{buno}
Gerhard Buntrock and Friedrich Otto (1998).
Growing context-sensitive languages and Church-Rosser languages.
{\em Inform. and Comput. \bf 141,} 1--36.
\item
\label{cn}
Yves Cochet and Maurice Nivat (1971).
Une g\'en\'eralization des ensembles de Dyck.
{\em Israel Journal of Mathematics \bf 9}, 389--395.
\item
\label{hartmanis}
Juris Hartmanis (1968).
On the complexity of one tape Turing machine
computations. {\em JACM \bf 15,} 325--339.
\item
\label{jl}
Tomasz Jurdzi\'nski and Krzysztof Lory\'s (2002).
Church-Rosser Languages vs. UCFL.
{\em Proc. 29th ICALP Symposium, Springer LNCS 2380}, 147--158.
\item
\label{jl07}
Tomasz Jurdzi\'nski and Krzysztof Lory\'s (2007).
Lower bound technique for length-reducing automata.
{\em Information and Computation \bf 205}, 1387--1412.
\item
\label{mcn}
Robert McNaughton (1999).
An insertion into the Chomsky Hierarchy?
{\em Jewels are forever, Contributions on Theoretical Computer
Science in Honor of Arto Salomaa,} 204--212.
\item
\label{mno}
Robert McNaughton, Paliath Narendran, and Friedrich Otto
(1988).  Church-Rosser Thue systems and formal languages.
{\em J. Association of Computing Machinery \bf 35}, 124--144.
\item
\label{naren}
P. Narendran (1984). {\em Church-Rosser and related Thue systems}.
Doctoral dissertation, Rensselaer Polytechnic Institute.
\item
\label{no}
Gundula Niemann and Friedrich Otto (2005).
The Church-Rosser languages are the deterministic variants
of the growing context-sensitive languages.
{\em Information and Computation \bf 197}, 1--21.
\item
\label{nivat}
Maurice Nivat (1970).  On some families of languages related to the
Dyck Languages. {\em Proc. 2nd ACM STOC,} 221--225. 
\item
\label{yu}
Sheng Yu (1989).
A pumping lemma for deterministic context-free languages.
{\em Information Processing Letters \bf 31,} 45--51.
%\item
%\label{od}
%Colm \'O D\'unlaing (1983).
%Infinite regular Thue systems.
%{\em Theoretical Computer Science \bf 25}, 171--192.
\end{enumerate}

\end{document}